\documentclass[onecolumn,aps,superscriptaddress,11pt,a4paper,pre]{article}
\usepackage[latin1]{inputenc}
\usepackage{amsmath}
\usepackage{amsfonts}
\usepackage{amssymb}
\usepackage{epsfig}
\usepackage{dcolumn}
\usepackage{xspace}

\begin{document}
\title{Effect of Feshbach Resonance on the Entropy Production  in the ultra cold Bosonic  Atoms   }
\author{M.N. Sinha Roy[1]\protect\footnote 
{corresponding author, Phone: 91-33-22411977; E-mail ID:msr$_{-}$presi@yahoo.com}, \\
{\it  Department of Physics,}\\
{Presidency College, 86/1 College street, Kolkata - 700 073, 
India}\\
\\
{}}
\date{\today}
\maketitle
\begin{abstract}
\noindent
The entropy of the coexisting  gas of ultra cold fermionic  atoms  and Bosonic molecular condensate confined in a
 magnetic trap has been  calculated from equation of motion  approaches. We have found that the entropy production
depends not only  to the Feshbach resonance and but also under certain limits to Rabi type oscillation and Bose Josephson junction type oscillation. 

 \end{abstract}

\vskip.1cm
PACS numbers: 03.75.Kk,67.85.Hj,67.85.Jk,67.85.De\\
Keywords: Entropy, Bosonic atoms, Condensates, B E C, superfluid .
\newpage
\section{Introduction}
The recent experimental observation of Bose - Einstein condensation (B E C) in a dilute gas of trapped atoms
\cite{mha} has generated much interest in the properties of this new state of matter. From the ensuing theoretical and experimental works \cite{don,ket,ang} it has been concluded that these condensed atomic gases behave very differently from the ideal non interacting gases, which yields the prospect of potentially displaying a rich phenomenology that include vortex states and Josephson like effect\cite{gbc}etc. A fascinating possibility is the observation  of new quantum phenomena on macroscopic scales, related with the superfluid nature of the condensate\cite{chin,rom}.\\
The G P E has been successfully applied to investigate the different properties  of trapped B E C in various approximations \cite{legg} and because of the nonlinear self-interaction, it could also induce chaotic behaviour in dynamical quantum observables \cite{sme,zap,yuk}.\\ 
In this paper we want to determine  the entropy of production in the ultra cold co - existing fermeonic atoms, Bosonic molecules  and the molecular condensate of Bosonic  atoms. The co existing ultra cold mixture of the atoms and molecular condensates  gases will be treated as two -component  fluid with negligible dissipation. A suitable form of Hamiltonian for the system has been considered for the calculation. This  mixture of the gaseous atomic component, uncondensed Bosonic molecules and the molecular condensate is close to transition temperature($ T_c $). The conserved variables in the system are : the energy, the  momentum, and the particle number densities. \\
\section{Formalism}
 We now treat the coexisting ultra cold fermionic atoms, uncondensed Bosonic molecules and the Bosonic molecular condensate near transition temperature $(T_c)$ as a two - component fluid  system. We use the equation of motion approach, from the knowledge of the total Hamiltonian of the system,  to determine the corresponding entropy and investigate how the entropy of the fluid depends on the atomic and molecular densities and the strength of the Feshbach interaction which is used to tune the interatomic attraction by variation of a magnetic field. This system also allows to study the cross over problem from BCS to BEC at the transition temperature. We assume that this fluid moves with velocity ($\vec v $)or rotates with angular velocity ($\vec{\omega}$)slowly within the harmonic trap. we shall consider the first case.  The fermionic atoms, uncondensed Bosonic molecules  may be treated as the normal component of the two - component fluid. \\
 
 The total Hamiltonian of the two - component fluid can be written as
 \begin{equation}
 {\cal H} = {\cal H}_{0}-\vec {P}.\vec v,
 \end{equation}
Here $\vec{P}$ is the total momentum  and ${\cal H}_0 =\sum_{\beta}^{\infty}{\frac{(p^\beta)^2}{2m_{\beta}}} + U$, with respect to rest  frame.
 The canonical partition function for this fluid, with N particles in a volume V is\cite{Landau}
 \begin{equation}
 Z_N(T,V,\vec v) = Tr e^{-\beta({\cal H} + \vec P.\vec v )}
 \end{equation}
 
 The thermodynamic potential,
 \begin{equation}
 F(T,V,N,\vec v) = -T ln {Z}_N = E - TS ,
 \end{equation}
 satisfies
 \begin{equation}
 dF = -S dT -p dV + \mu dN ,
 \end{equation}
 where the average of ${\cal H}_0$ is the internal energy $E$. $S$ and $T$ are the entropy and the temperature of the condensate. The chemical potential is $\mu$.
 It is now useful to introduce the grand potential,
 \begin{equation}
 {\cal A}(T,\mu,\vec v,V)= F -\mu N,
 \end{equation}
 
 which is a function of only one extensive variables V and satisfies 
 \begin{equation}
 d{\cal A}= S d T - p  V - N d \mu .
 \end{equation}
 The grand potential can be written as ${\cal A}=-V p(\mu,T,\vec v)$, where $ p=(\varepsilon-\eta \rho- T s - \vec g .\vec v)$. Here $\eta$ is the chemical potential per unit mass. Since T and $s = V^{-1} S$ are independent of $\vec v$, the above equation leads  to entropy equation
 \begin{equation}
 T d s = d \varepsilon - \eta d \rho - \vec{v}.d\vec{g},
 \end{equation}
 where the momentum density is $\vec g = \vec{v}\rho$, and $\rho = $ is the density of the condensate  and $\varepsilon(\vec r,t) = V^{-1}<{\cal H}_0>$,  the average energy density. The variations of all conserved variables have been included in the thermodynamic potentials. Therefore the entropy production equation becomes
 \begin{equation}
 T(\frac{\partial{s}}{\partial t} + \nabla.(\frac{{\vec j}}{T})) = - {\vec j}.(\frac{(\nabla T)}{T}) -(\eta +v^2)\frac{\partial{\rho}}{\partial t},
 \end{equation}
 where $\vec j(\vec r,t)$ represents the average energy current density. $\varepsilon(\vec r,t) $ and 
 $\vec j(\vec r,t)$ satisfy conservation law. Integrating above equation over the large volume subject to 
 boundary condition that $\vec j$ goes to zero at its outer surface, gives  the total 
 rate of entropy generation as
 \begin{equation}
 T(\frac{d S}{d t}) = \int d^3r [-{\vec j}.(\frac{(\nabla T)}{T})-(\eta + v^2).\frac{\partial{\rho}}{\partial t}].
 \end{equation}
 The density ($\rho$)of the system will  be written in terms of the atomic $(\rho_a)$ and the molecular $(\rho_m)$ densities. The population imbalance between atoms which are paired and those which are unpaired will be considered
 for the production of entropy.\\
 Now, we introduce the basic Hamiltonian for the fermionic atoms and Bosonic molecules for the determination of $\rho_a$ and $\rho_m$.  The operators($\psi_a$ and $\psi_m$)for atomic and molecular species,respectively, to be described by the given Hamiltonian \cite{rad,tim}
 \begin{eqnarray}
 H &=&\int d^3r{\psi}^{\dagger}_a(\vec{r})(\frac{-\hbar^2}{2m}\nabla^2)\psi_a(\vec{r})+
 \int d^3r\psi^{\dagger}_m(\vec{r})
 (\frac{-\hbar^2}{4m}\nabla^2 \nonumber +\epsilon)\psi_m(\vec{r})\nonumber\\
 &&+\frac{\lambda_a}{2}\int\psi^{\dagger}_a(\vec{r})\psi_a(\vec{r})
 \psi^{\dagger}_a(\vec{r})\psi_a(\vec{r})d^3r  \nonumber\\
 &&+\frac{\lambda_m}{2}\int\psi^{\dagger}_m(\vec{r})\psi_m(\vec{r})\psi^{\dagger}_m(\vec{r})
 \psi_m(\vec{r})d^3r 
 +\lambda\int\psi^{\dagger}_a(\vec{r})\psi^{\dagger}_m(\vec{r})\psi_a(\vec{r})
 \psi_m(\vec{r})d^3r  \nonumber\\
 &&+\frac{\alpha}{2}\int d^3r [\psi^{\dagger}_m(\vec{r})\psi_a(\vec{r})\psi_a(\vec{r})
 +\psi_m(\vec{r})\psi^{\dagger}_a(\vec{r})\psi^{\dagger}_a(\vec{r})].
 \end{eqnarray}
 In the above $\epsilon$ is the binding energy of the molecule. $\lambda_a$, $\lambda_m$ and
 $\lambda$ represent the strengths of the atom-atom, molecule-molecule and atom-molecule 
 interactions. The  interaction strength $\alpha$ characterizes the Feshbach  resonance \cite{fesh}
 term which leads to molecule formation and dissociation. The fluctuations of the fields are neglected and they are replaced by their expectation values which are classical 
 variables. Therefore the  operators $\psi_{a,m}(\vec r,t)$ are replaced by classical fields 
 $\phi_{a,m}(\vec r,t)$ and the corresponding Heisenberg equations of motion reduce 
 to coupled time dependent equations \cite{gross} as  
 \begin{equation}
 i\hbar\dot{\phi_a} = \frac{-\hbar^2}{2m}\nabla^2\phi_a + N_a \lambda_a|\phi_a|^2\phi_a +
 N_m\lambda|\phi_m|^2\phi_a + N_m^{1/2}\sqrt{2}\alpha \phi_m \phi^*_a,
 \end{equation}
 \begin{equation}
 i\hbar\dot{\phi_m} = (\frac{-\hbar^2}{4m}\nabla^2+\varepsilon)\phi_m+N_m\lambda_m|\phi_m|^2\phi_m+
 N_a\lambda|\phi_a|^2\phi_m + \frac{\alpha}{\sqrt{2}}\frac{N_a}{\sqrt{N_m}} \phi^2_a.
 \end{equation} \\
We begin by seeking a spatially uniform solution of the form
\begin{equation}
\phi_{a,m}=\frac{A_{a,m}(t)}{\sqrt{V}}e^{\frac{-i\mu_{a,m}t+\theta_{a,m}t}{\hbar}},
\end{equation}
where $\mu_{a,m }$are
the chemical potentials and the amplitudes $A_{a,m}(t)$ are in general function of time. Inserting $\phi_{a,m}$ in the above equations and equating real and imaginary parts, we led to
\begin{equation}
\dot{\theta_a}=\lambda_a\rho_a A_a^2+ \lambda\rho_m A_m^2 -\mu_a + \alpha A_m{\sqrt{\rho_m}}
cos[\frac{(2\mu_a-\mu_m)t+ 2\theta_a-\theta_m}{\hbar}],
\end{equation}
\begin{equation}
\hbar \dot{A_a} = \alpha A_a A_m {\sqrt{\rho_m}}sin[\frac{(2\mu_a-\mu_m)t + 2\theta_a -\theta_m }{\hbar}],
\end{equation}
 
\begin{equation}
 \dot{\theta_m} = \lambda_m\rho_m A_m^2 + \lambda\rho_a A_a^2 + \epsilon - \mu_m + 
 \frac{\alpha A_a^2}{2A_m}\frac{\rho_a}{\sqrt{\rho_m}}
 cos[\frac{(\mu_m - 2\mu_a)t+ 2\theta_m - 2\theta_a}{\hbar}]
 \end{equation}\\
  \begin{equation}
 \hbar \dot{A_m} = \frac{\alpha}{2} A_a^2 \frac{\rho_a}{\sqrt{\rho_m}}sin[\frac{(2\mu_m - 2\mu_a)t + \theta_m - 2\theta_a }{\hbar}],
 \end{equation}\\
 
 In the above  $\rho_a = \frac{N_a}{V}$ and $\rho_m = \frac{N_m}{V} $. Here $ N_a$ and $N_m$ are the 
 actual number of atoms and molecules at any time.\\  
 In the  hydrodynamic approximation with $\rho_a = A_a^2$ and $\rho_m = A_m^2$ and 
$\phi = (2\mu_a-\mu_m)t + 2\theta_a - \theta_m$, the above equations become
  \begin{equation}
\dot{\rho_a}= \frac{\alpha}{\hbar}\frac{\rho_a\rho_m}{2}sin{\phi},
 \end{equation}
 
  \begin{equation}
 \dot{ \rho_m} = -\frac{\alpha}{\hbar}\rho_a^2 sin{\phi},
  \end{equation}
 \begin{equation}
 \phi = -\epsilon + (2\lambda_a - \lambda)\rho_a^2 + (2\lambda - \lambda_m) \rho_m^2 +
 \frac{\alpha}{\hbar}(2\rho_m - \frac{\rho_a^2)}{2\rho_m})cos{\phi}.
 \end{equation}
 We note immediately that $2\rho_a^2+\rho_m^2$ is independent of time and call this $c_0$. we now write in terms 
 of the variable $ z = \rho_a^2 - 4\rho_m^2$ and $\phi$, to write
 \begin{equation}
 \dot{z} = \frac{\sqrt{2}\alpha}{3\hbar}(8 c + z)( c - z)^{1/2}sin{\phi}
  \end{equation}
 
 \begin{equation}
 \dot{\phi} = \Delta_1 - \Lambda z - \frac{3\alpha}{2\sqrt{2}\hbar} \frac{z}{\sqrt{(c -z)}} cos{\phi}
 \end{equation}
where\\
\begin{equation}
 \dot{\Delta}_1 = -\epsilon + \frac{c}{9}[16 \lambda _a - 4 \lambda - 2 \lambda_m]
 \end{equation}
 and\\
 \begin{equation}
 \Lambda = \frac{-1}{9}(2\lambda_a + 2\lambda_m - 5 \lambda)
 \end{equation} 
 For $z << 1$ and $\phi <<1$ , we have a simple harmonic motion with frequency 
 $\omega^2=(\Lambda + \frac{3\alpha}{2\sqrt{2}\hbar})\frac{8\sqrt{2}}{3\hbar}c\alpha$. The linearisation above means 
 $\rho_a \simeq 2\rho_m$. Thus  the variable z is the population  imbalance  between atoms which are paired and those 
 which are unpaired.
 If the interaction  among the atoms and the molecules are neglected  and only the Feshbach resonance 
 (i.e. $\alpha $) term is considered then we have a simple harmonic motion  with frequency $\omega = \frac{2\alpha}{\hbar}$
 between the atoms and the molecular condensate. This is pure Rabi oscillation. With the interaction present, the oscillations, according to Smerzi et al.\cite{sme} and Zapata et al. \cite{zap}, are a Bose Josephson junction oscillation
which are responsible for the oscillations. An interesting limit is when z is small but $\phi$ can be arbitrary. In this case, equations for z and $\phi$ reduce to 
  \begin{equation}
  \dot{z} = \frac{8\sqrt{2}\alpha c^{3/2}}{3\hbar}sin{\phi}
  \end{equation}
  \begin{equation}
  \dot{\phi} =   {\Delta}_1 - \Lambda z -\frac{3\alpha z}{2\sqrt{2}c\hbar}cos{\phi} 
  \end{equation}
 
 If ${\Delta}_1$ happens to be larger than $\Lambda z + \frac{3\alpha z}{2\sqrt{2}c\hbar}cos{\phi}$,then 
 $\dot{\phi} = {\Delta}_1$ and $\phi = {\Delta}_1 t$. It is now clear that z oscillates with frequency 
  ${\Delta}_1$ i.e. the entropy production exhibits oscillations.  \\ 
 It should be noted that in the limit of $\frac{\alpha c}{\hbar} = \beta >> 1$, the equations (25) and (26)
 admit an exact answer including the nonlinear interaction. In this case with z/c = y , we can write 
 \begin{equation}
 \dot{y} = \frac{8\sqrt{2} \beta}{3}(1 + y/8)(1 - y)^{1/2} sin{\phi}
 \end{equation}
 
 \begin{equation}
 \dot{\phi} = -\frac{3\beta}{2\sqrt 2} \frac{y}{\sqrt{1-y}} cos {\phi}
 \end{equation}
 The above equations lead to 
 \begin{equation}
 \frac{dy}{dx}  =  -\frac{32}{9} \frac{(1+y/8)}{(1-y)}tan{\phi}
 \end{equation}\\
 which integrate to 
 \begin{equation}
 \frac{1}{4} ln[\frac{(1 - y)}{1 - y/8}] = ln|sec{\phi|} + I_0
 \end{equation}
 
 where $I_0$ is a constant of integration. We finally observe that $\Lambda$ can become negative and if the 
 atom - atom and molecule - molecule scattering dominates the auto - molecule scattering, then it is possible for the oscillation to be quenched. In conclusion, we have seen that a gas of ultra cold Bosonic atoms with a resonance term can allow the formation and dissociation of molecules and is capable of showing oscillations which are Rabi type oscillations and Josephson like in certain limits and this suggests that the entropy production  in the system is possible. This is oscillatory, dependent on the atomic and molecular densities and also on the  Feshbach resonance term, even when the temperature  the system becomes constant or the system is thermally insulated. \\ 
  Now suitably controlling the magnetic field, the temperature of the  cold atoms decreases and the Feshbach 
 interaction induces the formation of the molecular condensate. It is interesting to note that the rate of entropy production and its  frequency of oscillation  can be manipulated by varying the strength of Feshbach resonance with the help of external magnetic field.

\end{document}